# Fermi surface of Kagome metal CsCr$_3$Sb$_5$ observed by laser photoemission microscopy

Hayate Kunitsu[1], Iori Ishiguro[1], Natsuki Mitsuishi[1], Shunsuke Tsuda[2], Koichiro Yaji[2,3], Zehao Wang[4,5], Pengcheng Dai[4,5], Yoichi Yamakawa[1], Hiroshi Kontani[1], Takahiro Shimojima[1]

[1] *Department of Physics, Nagoya University, Furo-cho, Nagoya 464-8602, Japan*
[2] *Center for Basic Research on Materials, National Institute for Materials Science, Tsukuba 305-0047, Japan*
[3] *Unprecedented-scale Data Analytics Center, Tohoku University, Sendai 980-8578, Japan*
[4] *Department of Physics and Astronomy, Rice University, Houston 77005, USA*
[5] *Rice Laboratory for Emergent Magnetic Materials and Smalley-Curl Institute, Rice University, Houston, Texas 77005, USA*

**We investigated the Fermi surface (FS) in the paramagnetic state of Kagome metal CsCr$_3$Sb$_5$ by employing a laser photoemission microscopy. We found a circular FS and two hexagonal FSs around the Brillouin zone (BZ) center. Polarization-dependent measurements further enable us to detect small FS pockets at the BZ boundary. According to the density functional theory calculations, the orbital characters of the FSs were determined from their shape and orientations. We found that the size of the FS is strongly modified for the $d_{xz}$ orbital, suggesting the orbital-dependent correlation effect. These results provide an electronic basis for exploring the interplay of antiferromagnetic/charge density wave order and possible unconventional superconductivity in this compound.**

Kagome materials exhibit novel physical properties such as spin liquids,[1–4] topological quantum phases[5–9] and unconventional superconductivity[10–13] arising from an interplay between topology, electron correlation and geometrical frustration. These phenomena arise from the characteristic electronic features of the Kagome lattice, i.e.

Dirac cones, van Hove singularities (vHS), and Flat bands (FBs), tuned at the Fermi level ($E_F$). $A$V$_3$Sb$_5$ family ($A$ = K, Rb, and Cs) exhibits superconductivity in the nonmagnetic Kagome lattice with weak electronic correlation. The X-ray diffraction (XRD) studies reported the 2 × 2 × 2 superstructure, suggesting the charge density wave (CDW) formation,[14] which might be chiral and associated with the time-reversal symmetry breaking.[15] Previous angle-resolved photoemission spectroscopy (ARPES) reported that the CDW order originates in the nesting between vHS near $E_F$.[16]

In contrast, recently discovered Kagome metal CsCr$_3$Sb$_5$ exhibits magnetism, charge order, and strong correlation[17]. Anomalous behaviors at 55 K in transport and thermodynamic properties suggested multiple orders in both charges and spins. Nuclear magnetic resonance measurements reported antiferromagnetic order. The observation of the 4 × 1 superlattices from the XRD measurements indicates the CDW formation with the rotational symmetry breaking. Superconductivity emerges at the transition temperature ($T_c$) of 6.4 K at 4.2 GPa. Unconventional electron pairing mechanism has been considered for this compound because of the non-Fermi liquid behavior and the upper critical field exceeding the Pauli limit around the maximum $T_c$.[17]

Characteristic electronic structure in CsCr$_3$Sb$_5$ is the FB-like feature at the binding energy of 80 meV observed by ARPES.[18–20] The density functional theory (DFT) calculations also suggested the presence of the FB at ~300 meV above $E_F$,[21] which may dominate the correlated electronic properties of this compound. Furthermore, theoretical studies proposed the importance of the FS nesting condition which enhances the antiferromagnetic fluctuations leading to the unconventional $s\pm$ or $d$ wave superconductivity.[21] The FS-related itinerant magnetism was, on the other hand, proposed to be important for the formation of the antiferromagnetic order in CsCr$_3$Sb$_5$.[22] While the extensive ARPES studies reported the band structure of CsCr$_3$Sb$_5$,[18–20,23] the complete determination of the FSs has not been achieved so far. It has been shown that the electronic band dispersions in CsCr$_3$Sb$_5$ are much broader than $A$V$_3$Sb$_5$ family,[19] possibly due to the strong correlation effect[20] and/or structural inhomogeneity.[24] For understanding the origin of the DW phase transitions and superconductivity in CsCr$_3$Sb$_5$, the experimental determination of the FSs has been highly demanded.

In this study, we report the FS in the paramagnetic state of CsCr$_3$Sb$_5$ by employing a laser photoemission microscopy, with a detecting area of 10 × 30 μm$^2$. We succeeded in observing all the FSs, i.e. three concentric FSs around the Brillouin zone (BZ) center and the small FS pockets at the BZ boundary, from a small area of the sample surface. We further determined the orbital characters of these FSs from their shape and orientations, according to the DFT calculations. The number of the FS is consistent with the DFT calculations, while the FS sheets derived from $d_{xz}$ orbital showed a large difference in size from the calculations.

The high-quality single crystals of CsCr$_3$Sb$_5$ were synthesized by self-flux method as reported in elsewhere.[18] The sample was mounted on the tantalum plate with silver paste and cleaved *in situ* at room temperature under the vacuum of 1.0 × 10$^{-10}$ Torr. Imaging-type spin-resolved photoemission microscopy (iSPEM)[25–27] with a 10.9 eV laser at NIMS was employed to visualize the FS in entire first BZ. The polarization-dependent measurements were performed by employing left- and right-circularly polarized (LCP and RCP) lasers.

CsCr$_3$Sb$_5$ crystalizes in a hexagonal lattice with the space group *P*6/*mmm* [Figs. 1(a) and 1(b)]. The DFT calculation indicates one electron band (Sb $p_z$) and two hole bands (Cr $d_{xz}$ and Cr $d_{x2-y2}$) crossing $E_F$ around Γ(A) point [Fig. 1(c)]. These bands tend to form nearly degenerated FS sheets around $k_z$ = 0 as shown in Fig. 1(d). The electron band at M(L) point forms the FS pockets showing a large $k_z$ dependence in size [Fig. 1(e)]. In the laser photoemission microscopy data, we use the labeling of $\bar{\Gamma}$, $\bar{M}$ and $\bar{K}$ in the two-dimensional BZ projected onto (001) surface.

First, we show the FSs of CsCr$_3$Sb$_5$ at 70 K (paramagnetic state) in Figs. 2(a) and 2(b), obtained by LCP and RCP lasers, respectively. Overall FS shapes are quite similar for both measurements, while the location of the high intensity exhibits a polarization dependence. One can recognize the circular FS around $\bar{\Gamma}$ point and high-intensity spots around $\bar{M}$ point. In order to clarify the weak intensity laying between BZ center and boundary, we performed the curvature analysis[28] for the raw FS data. The disconnected portions of the FSs centered at $\bar{\Gamma}$ are emphasized in Figs. 2(c) and 2(d), which are consistently observed in the peak-plot analysis of the raw FS data as highlighted by the gray curves in Figs. 2(e) and 2(f). Here we overlay the markers obtained from the peak plot analysis for LCP and RCP data as shown in Fig. 2(g). The disconnected portions of the FSs [gray curves in Fig. 2(g)] can be assigned to the upper (lower) part of the hexagonal FS with the corners directed to $\overline{\text{M}}$ ($\overline{\text{K}}$) point. We note that in the present experimental geometry, the photoelectron distribution in the momentum space should be anisotropic and some portions of the FSs cannot be detected. As summarized in Fig. 2(h), the presence of the inner ($\alpha$), middle ($\beta$) and outer ($\gamma$) FS sheets were confirmed around $\overline{\Gamma}$ point.

The polarization dependence of the FS is rather clear around $\overline{\text{M}}$ point [Fig. 2(g)]. The LCP and RCP lasers seem to detect the inner or outer part of the elliptical FS pockets ($\delta$) separated by the BZ boundary [black lines in Fig. 2(g)]. In order to investigate this polarization dependence, we show the line profiles of the raw FS data [Figs. 2(a) and (b)]. Figure 3(a) exhibits the line profiles along $\overline{\Gamma}$-$\overline{\text{M}}$ obtained by the LCP and RCP lasers at azimuthal angles ($\theta$) of 60 degrees and 120 degrees. While the line profiles at $\theta = 60º$ (black) are almost identical, those for $\theta = 120º$ (gray) show clear polarization dependence especially in the peaks for $\delta$ and $\beta$ FSs. The inverted open triangles in Fig. 3(a) highlight the difference in the peak positions for the $\delta$ FS, reflecting two Fermi momentum ($k_F$) for the left and right part of the elliptical FS pocket around $\overline{\text{M}}$ point. On the other hand, the inverted filled triangles on the line profiles at $\theta = 120º$ suggest the presence of the peak for the $\beta$ FS in the LCP data which is absent in the RCP data. The pair of the peaks at ±0.2 Å$^{-1}$ corresponds to the $\alpha$ FS. Finally, the hump structure around +0.55 Å$^{-1}$ can be assigned to the $\gamma$ FS. Then, the line profile taken by LCP at $\theta = 120º$ is well reproduced by the fitting function composed of six Lorentz functions (red dotted curves) with a background (black dotted line) in Fig. 3(b), demonstrating the complete assignment of the FSs in CsCr$_3$Sb$_5$.

Here we discuss the orbital character for each FS sheet. We found that the observed FSs around $\overline{\Gamma}$ show different shapes and orientations i.e. circular $\alpha$ FS, hexagonal $\beta$ FS with the corners directed to K point, and hexagonal $\gamma$ FS with the corners directed to M point. According to the calculated FSs at $k_z = \pi$ [Fig. 1(e)], the shape and orientation of the inner hexagonal FS composed of Cr $d_{xz}$ orbital (blue) is identical to those of the $\gamma$ FS. The middle circular FS formed by the electron band of Sb $p_z$ orbital (green) corresponds to the $\alpha$ FS. Then, the outer hexagonal FS of Cr $d_{x2-y2}$ orbital (orange) is assigned to the β FS. Finally, the δ FS pockets at $\bar{\text{M}}$ point are determined to have Cr $d_{xz}$ orbital character (blue). We consider that the experimental $k_z$ value is closer to π rather than 0 according to the observation of the well-separated three FSs around $\bar{\Gamma}$.

The total number of the FSs in the present data is consistent with the DFT calculation [Figs. 1(d) and 1(e)]. However, the size of the γ (δ) FS is much larger (smaller) than the calculated one. A possible explanation for these quantitative disagreements might be the orbital-selective band renormalization.[29] In this scenario, the size and shape of the FS should be modified by the correlation effect, which is most prominent for the $d_{xz}$ orbital. For example, the correlated FS pockets at M point ($d_{xz}$) are smaller than those without correlations. In our data, both γ and δ FSs were suggested to have $d_{xz}$ orbital character. The disagreement in size of these FSs seems to be consistent with this scenario. The present results suggest the importance of electronic correlations for properly understanding the paramagnetic electronic states of $CsCr_3Sb_5$.

In conclusion, we investigated the FS in the nonmagnetic state of $CsCr_3Sb_5$ using the iSPEM. We found three concentric FSs around $\bar{\Gamma}$ point and small FS pockets around $\bar{\text{M}}$ point. According to the DFT calculations, the orbital characters of these FSs were determined from their shape and orientations. The size of the FS sheets formed by $d_{xz}$ orbital exhibits a large deviation from the DFT calculations, supporting the orbital-selective band renormalization. These results provide an important insight for exploring the antiferromagnetic/charge density wave order and superconductivity in $CsCr_3Sb_5$.

**Acknowledgement**

The authors thank F. Arai for the technical support of iSPEM measurements. The present work was partially supported by JST CREST, Japan (Grant No JPMJCR2435), the Japan Society for the Promotion of Science KAKENHI (Grant Nos. 24K01352, 24K17591, 25K07195), and the Innovative Science and Technology Initiative for Security Grant Number JPJ004596, ATLA, Japan. The single-crystal synthesis and characterization at

Rice were supported by the U.S. DOE, BES under Grant Nos. DESC0012311 and DE-SC0026179 (P.D.)

**References**

1. L. Balents, *Nature* **464**, 199–208 (2010).

2. S. Yan, D.A. Huse, and S.R. White, *Science* **332**, 1173–1176 (2011).

3. T.-H. Han, J.S. Helton, S. Chu, D.G. Nocera, J.A. Rodriguez-Rivera, C. Broholm, and Y.S. Lee, *Nature* **492**, 406–410 (2012).

4. M. Fu, T. Imai, T.-H. Han, and Y.S. Lee, *Science* **350**, 655–658 (2015).

5. E. Liu, Y. Sun, N. Kumar, L. Muechler, A. Sun, L. Jiao, S.-Y. Yang, D. Liu, A. Liang, Q. Xu, J. Kroder, V. Süß, H. Borrmann, C. Shekhar, Z. Wang, C. Xi, W. Wang, W. Schnelle, S. Wirth, Y. Chen, S.T.B. Goennenwein, and C. Felser, *Nat. Phys.* **14**, 1125–1131 (2018).

6. D.F. Liu, A.J. Liang, E.K. Liu, Q.N. Xu, Y.W. Li, C. Chen, D. Pei, W.J. Shi, S.K. Mo, P. Dudin, T. Kim, C. Cacho, G. Li, Y. Sun, L.X. Yang, Z.K. Liu, S.S.P. Parkin, C. Felser, and Y.L. Chen, *Science* **365**, 1282–1285 (2019).

7. N. Morali, R. Batabyal, P.K. Nag, E. Liu, Q. Xu, Y. Sun, B. Yan, C. Felser, N. Avraham, and H. Beidenkopf, *Science* **365**, 1286–1291 (2019).

8. L. Ye, M. Kang, J. Liu, F. Von Cube, C.R. Wicker, T. Suzuki, C. Jozwiak, A. Bostwick, E. Rotenberg, D.C. Bell, L. Fu, R. Comin, and J.G. Checkelsky, *Nature* **555**, 638–642 (2018).

9. X. Teng, L. Chen, F. Ye, E. Rosenberg, Z. Liu, J.-X. Yin, Y.-X. Jiang, J.S. Oh, M.Z. Hasan, K.J. Neubauer, B. Gao, Y. Xie, M. Hashimoto, D. Lu, C. Jozwiak, A.

Bostwick, E. Rotenberg, R.J. Birgeneau, J.-H. Chu, M. Yi, and P. Dai, *Nature* **609**, 490–495 (2022).

10. W.-H. Ko, P.A. Lee, and X.-G. Wen, *Phys. Rev. B* **79**, 214502 (2009).

11. M.L. Kiesel, and R. Thomale, *Phys. Rev. B* **86**, 121105 (2012).

12. M.L. Kiesel, C. Platt, and R. Thomale, *Phys. Rev. Lett.* **110**, 126405 (2013).

13. W.-S. Wang, Z.-Z. Li, Y.-Y. Xiang, and Q.-H. Wang, *Phys. Rev. B* **87**, 115135 (2013).

14. H. Li, T.T. Zhang, T. Yilmaz, Y.Y. Pai, C.E. Marvinney, A. Said, Q.W. Yin, C.S. Gong, Z.J. Tu, E. Vescovo, C.S. Nelson, R.G. Moore, S. Murakami, H.C. Lei, H.N. Lee, B.J. Lawrie, and H. Miao, *Phys. Rev. X* **11**, 031050 (2021).

15. T. Asaba, A. Onishi, Y. Kageyama, T. Kiyosue, K. Ohtsuka, S. Suetsugu, Y. Kohsaka, T. Gaggl, Y. Kasahara, H. Murayama, K. Hashimoto, R. Tazai, H. Kontani, B.R. Ortiz, S.D. Wilson, Q. Li, H.-H. Wen, T. Shibauchi, and Y. Matsuda, *Nat. Phys.* **20**, 40–46 (2024).

16. K. Nakayama, Y. Li, T. Kato, M. Liu, Z. Wang, T. Takahashi, Y. Yao, and T. Sato, *Phys. Rev. B* **104**, L161112 (2021).

17. Y. Liu, Z.-Y. Liu, J.-K. Bao, P.-T. Yang, L.-W. Ji, S.-Q. Wu, Q.-X. Shen, J. Luo, J. Yang, J.-Y. Liu, C.-C. Xu, W.-Z. Yang, W.-L. Chai, J.-Y. Lu, C.-C. Liu, B.-S. Wang, H. Jiang, Q. Tao, Z. Ren, X.-F. Xu, C. Cao, Z.-A. Xu, R. Zhou, J.-G. Cheng, and G.-H. Cao, *Nature* **632**, 1032–1037 (2024).

18. Z. Wang, Y. Guo, H.-Y. Huang, F. Xie, Y. Huang, B. Gao, J.S. Oh, H. Wu, J. Okamoto, G. Channagowdra, C.-T. Chen, F. Ye, X. Lu, Z. Liu, Z. Ren, Y. Fang, Y. Wang, A. Biswas, Y. Zhang, Z. Yue, C. Hu, C. Jozwiak, A. Bostwick, E. Rotenberg, M. Hashimoto, D. Lu, J. Kono, J.-H. Chu, B.I. Yakobson, R.J. Birgeneau, G.-H. Cao, A. Fujimori, D.-J. Huang, Q. Si, M. Yi, and P. Dai, *Nat. Commun.* **16**, 7573 (2025).

19. S. Peng, Y. Han, Y. Li, J. Shen, Y. Miao, Y. Luo, L. Huai, Z. Ou, Y. Chen, D. Hu, H. Li, Z. Xiang, Z. Liu, D. Shen, M. Hashimoto, D. Lu, X. Li, Z. Qiao, Z. Wang, and J. He, *Sci. China Phys. Mech. Astron.* **69**, 217412 (2026).

20. Y. Li, Y. Liu, X. Du, S. Wu, W. Zhao, K. Zhai, Y. Hu, S. Zhang, H. Chen, J. Liu, Y. Yang, C. Peng, M. Hashimoto, D. Lu, Z. Liu, Y. Wang, Y. Chen, G. Cao, and L. Yang, *Nat. Commun.* **16**, 3229 (2025).

21. S. Wu, C. Xu, X. Wang, H.-Q. Lin, C. Cao, and G.-H. Cao, *Nat. Commun.* **16**, 1375 (2025).

22. C. Xu, S. Wu, G.-X. Zhi, G. Cao, J. Dai, C. Cao, X. Wang, and H.-Q. Lin, *Nat. Commun.* **16**, 3114 (2025).

23. Y. Li, P. Li, T. Miao, R. Xu, Y. Cai, N. Cai, B. Liang, H. Gao, H. Xiao, Y. Jiang, J. Cao, F. Zhu, H. Wang, J. Xie, J. Li, Z. Liu, C. Chen, Y. Zhang, X.J. Zhou, D. Zhong, H. Wang, J. Huang, and D. Guo, Preprint at 10.48550/arXiv.2510.12888 (2025).

24. T. Kato, Y. Li, K. Nakayama, Z. Wang, S. Souma, M. Kitamura, K. Horiba, H. Kumigashira, T. Takahashi, and T. Sato, *Phys. Rev. B* **106**, L121112 (2022).

25. K. Yaji and S. Tsuda, *e-J. Surf. Sci. Nanotechnol.* **22**, 46 (2024).

26. K. Yaji and S. Tsuda, *Sci. Technol. Adv. Mater. Methods* **4**, 2328206 (2024).

27. S. Tsuda and K. Yaji, *e-J. Surf. Sci. Nanotechnol.* **22**, 170 (2024).

28. P. Zhang, P. Richard, T. Qian, Y.-M. Xu, X. Dai, and H. Ding, *Rev. Sci. Instrum.* **82**, 043712 (2011).

29. F. Xie, Y. Fang, Y. Li, Y. Huang, L. Chen, C. Setty, S. Sur, B. Yakobson, R. Valentí, and Q. Si, *Phys. Rev. Res.* **7**, L022061 (2025).

**Fig. 1**

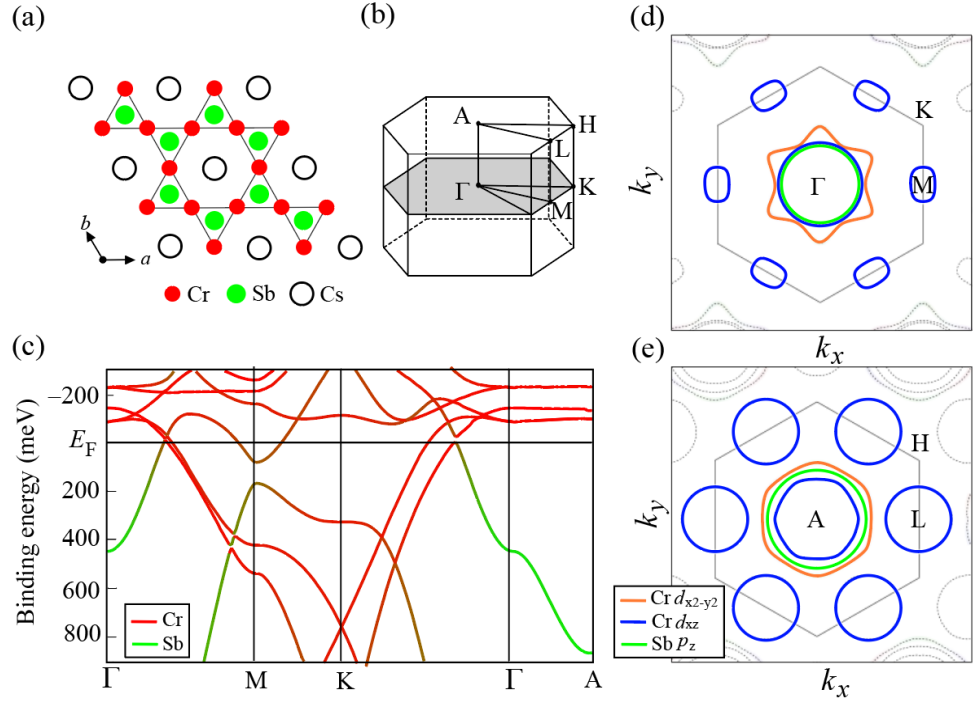

**Fig. 1.** Calculated band structure and Fermi surfaces of $CsCr_3Sb_5$. (a) Top view of the crystal structure of $CsCr_3Sb_5$. The red, green and white circles represent the Cr, Sb and Cs atoms, respectively. (b) Three-dimensional Brillouin zone. (c) The band structure of $CsCr_3Sb_5$ calculated by the density functional theory. The red and green curves represent the band dispersions of Cr $d$ and Sb $p$ orbital character, respectively. (d, e) The calculated Fermi surfaces of $CsCr_3Sb_5$ at $k_z = 0$ and $\pi$, respectively. Orange, blue and green curves represent the Fermi surface composed of Cr $d_{x2-y2}$, Cr $d_{xz}$ and Sb $p_z$ orbital, respectively.

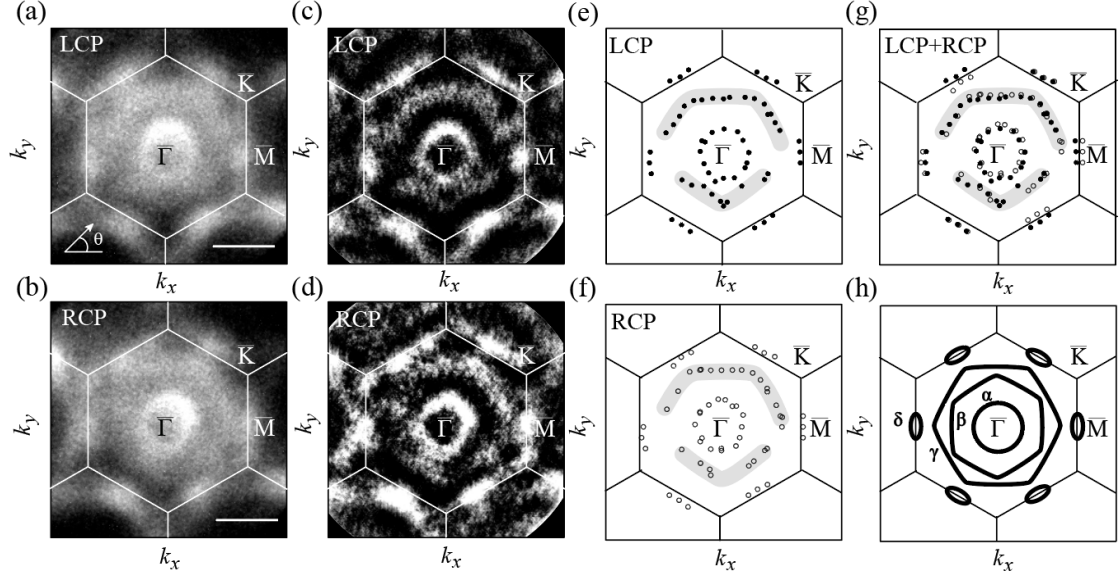

**Fig. 2.** Fermi surface of CsCr$_3$Sb$_5$ observed by laser photoemission microscopy. (a, b) Fermi surface mapping of CsCr$_3$Sb$_5$ measured with LCP and RCP laser, respectively. White lines represent the Brillouin zone boundaries. The definition of the azimuthal angle and scale bar corresponding to 0.5 Å$^{-1}$ are shown. (c, d) Curvature analysis of the data in (a) and (b), respectively. (e, f) The peak plots of the data in (a) and (b), respectively. Thick gray curves indicate the portions of the Fermi surfaces β and γ. (g) Superimposed peak plots in (e) and (f). (h) Summary of the Fermi surface observations.

**Fig. 3**

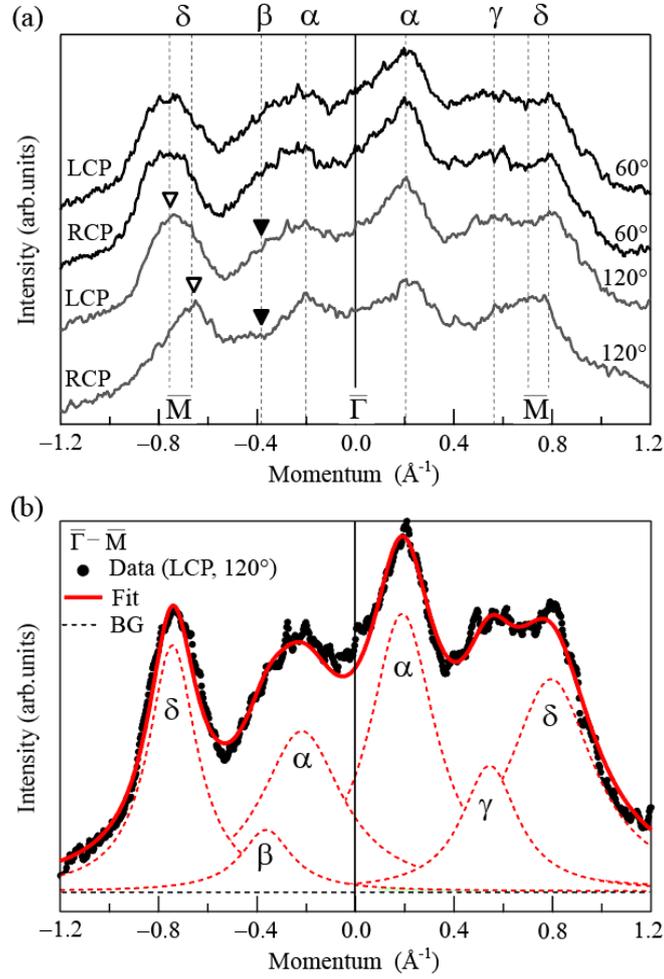

**Fig. 3.** Intensity profiles of the Fermi surface mapping data. (a) Intensity profiles along $\bar{\Gamma}$–$\bar{M}$ line at the azimuthal angles of 60° (black) and 120° (gray) measured by LCP and RCP laser. The inverted open triangles highlight the difference in $k_F$ for the δ band, reflecting two $k_F$ values for the left and right side of the elliptical FS pocket at $\bar{M}$ point. The inverted filled triangles indicate the presence (LCP) and absence (RCP) of the intensity for the β FS. The dotted lines represent the $k_F$ for each FS. (b) Fitting analysis for the intensity profile at the azimuthal angles of 120° taken by LCP laser. Red curve represents the fitting function composed of six Lorentz functions (red dotted curves) and a constant background (black dotted line).